\documentclass[aps,pra,twocolumn,longbibliography,showpacs,10pt,floatfix]{revtex4-1}
\usepackage{graphicx}
\usepackage{epstopdf}
\usepackage[within=none]{newfloat}
\DeclareFloatingEnvironment[name=Video,placement=htbp]{movie}

\begin{document}
\title{Fracture size effects in nanoscale materials: the case of graphene}
\author{Alessandro Luigi Sellerio}
\author{Alessandro Taloni}
\affiliation{Center for Complexity and Biosystems, Department of Physics,  University of Milano, via Celoria 16, 20133 Milano, Italy}
\affiliation{CNR - Consiglio Nazionale delle Ricerche,  Istituto per l'Energetica e le Interfasi, Via Roberto Cozzi 53, 20125 Milano, Italy}
\author{Stefano Zapperi}
\email{stefano.zapperi@unimi.it}
\affiliation{Center for Complexity and Biosystems, Department of Physics,  University of Milano, via Celoria 16, 20133 Milano, Italy}
\affiliation{CNR - Consiglio Nazionale delle Ricerche,  Istituto per l'Energetica e le Interfasi, Via Roberto Cozzi 53, 20125 Milano, Italy}
\affiliation{Institute for Scientific Interchange Foundation, Via Alassio 11/C, 10126 Torino, Italy}
\affiliation{Department of Applied Physics, 
Aalto University, P.O. Box 14100, FIN-00076 Aalto, Espoo, Finland}



\begin{abstract}
Nanoscale materials display enhanced strength and toughness but also larger fluctuations and more pronounced
size effects with respect to their macroscopic counterparts. Here we study the system size-dependence of the failure strength distribution of a monolayer graphene  sheet with a small concentration of vacancies by molecular dynamics simulations. We simulate sheets of varying size encompassing more than three decades and systematically study their deformation as a function of disorder, temperature and loading rate. We generalize the weakest-link theory of fracture size effects to rate and temperature dependent failure and find quantitative agreement with
the simulations. Our numerical and theoretical results explain the crossover 
of the fracture strength distribution between a thermal and rate-dependent regime and a
disorder-dominated regime described by extreme value theory.
\end{abstract}

\maketitle

Nanomaterials have remarkable mechanical properties, such as enhanced strength and toughness \cite{mielke2007,greer2011}, but display considerable size effects and sample-to-sample fluctuations, which represent an issue for engineering applications.  Our current understanding of fracture size-effects in macroscopic disordered media relies on extreme value theory which relates the strength to the statistics of the weakest region in the sample \cite{gumbel,weibull39}. While the theory does not consider the effect of stress concentrations and crack interactions, numerical models for the failure of elastic networks with disorder show that an extreme value distribution describes failure at large enough scales, although the form usually deviates from the standard Weibull distribution \cite{ray1985,duxbury86,manzato2012,bertalan2014}. Understanding size effects in nanomaterials is still an intriguing open issue also because of the presence of rate-dependent thermal effects that would invalidate the weakest-link  hypothesis \cite{vanel2009}. Yet, the Weibull distribution is commonly used to fit experimental data in carbon based nanomaterials \cite{barber2005}, although the tensile strength is observed to depend on the strain rate \cite{sun2012}.

Testing fracture properties of graphene is quite challenging due to the difficulty in applying high tensile stresses in a controlled fashion on nanoscale objects \cite{Lee2008,Kim2012,Zhang2014}. Therefore numerical simulations represents a viable alternative to understand the size dependence of its mechanical behavior \cite{belytschko2002atomistic,lu05,Hartmann2013,Moura2013,Xu2013}. Numerical simulations of defected carbon nanutubes suggest that failure is described by the Weibull distribution in quasistatic, zero-temperature conditions \cite{Yang2007}. Finite temperature molecular dynamics simulations reveal, however, that the average tensile strength of nanotubes \cite{zhao2010} and graphene \cite{Zhao2009,Ansari2012274,Xu2013} depends on temperature and loading rate. Despite these insightful results, a comprehensive theory describing the size dependent fracture strength distribution of carbon nanomaterials, elucidating the role of thermal fluctuations and strain rate, is still lacking.
 
Here we perform large scale molecular dynamics simulations of the deformation and failure of defected monolayer graphene sheets for a wide range of sample sizes, vacancy concentration, temperature and strain rate. To explain the observed temperature and rate dependence of the tensile strength distribution, we generalize extreme value theory to the case of thermally activated rate dependent fracture. The resulting theory is shown to be in excellent agreement with our simulations and provides a general framework to explain rate-dependent  thermal effects in the failure of disordered nanomaterials. Based on our theory, we 
derive a simple criterion that allows to assess the relative importance of structural disorder and thermal fluctuations in determining failure. Using this rule, one can readily show that the failure of nanoscale samples is more prone to thermal
induced failure, while the fracture macroscopic samples are more likely to be ruled by quenched disorder. This confirms previous results showing that in the limit of very large samples failure is ruled by extreme value statistics (although not necessarily by the Weibull law) \cite{shekhawat2013,bertalan2014}. 

The paper is organized as follows. In section \ref{sec:model} we describe the molecular dynamics simulation model and in section \ref{sec:simulations} discuss the numerical results. The theory is described in details in section
\ref{sec:theory} where we also compare its prediction with experiments. Section \ref{sec:discussion} discusses
the general implications of our work to understand size effects in materials at different scales. Appendix \ref{sec:potential} provides details on the choice of interatomic potential and appendix \ref{sec:fit} discusses the fitting method.

\section{Model \label{sec:model}}
We perform numerical simulations of the deformation and failure of defected monolayer graphene  using the LAMMPS molecular dynamics simulator package~\cite{Plimpton1995}. The carbon-carbon atom interaction is modeled with  the ``Adaptive Intermolecular REactive Bond Order'' (AIREBO) potential~\cite{Stuart2000}.  In order to simulate a realistic bond failure behavior, the shortest-scale adaptive cutoff of the AIREBO potential has to be fine-tuned \cite{belytschko2002atomistic,Zhao2009}, as detailed in appendix \ref{sec:potential}. The simulated system consists of single layer, monocrystalline graphene sheets, composed of a variable number \(N\) of atoms: \(N\) varies from approximately $10^3$ to $50\times 10^3$ atoms. The sheets are prepared by placing the atoms on a hexagonal lattice; the characteristic lattice length scale $\lambda = 1.42 ~\AA$ is chosen so that the system is initially in an equilibrium configuration. The sheets have  an almost square shape lying on the XY coordinate plane; their lateral size depends on  $N$ and varies between 50~and $360 ~\AA$ (5~and 36~nm). When placing defects on the sheets, a fixed fraction of atoms is randomly removed; this corresponds to vacancy concentrations $P = 0.1$, 0.2~and 0.5\%. While the graphene layer is essentially 2D, the atom positions are integrated in all the three spatial directions; also, the layers have no periodic boundary conditions.

The simulations are performed by  stretching  the samples  along  the X coordinate axis, corresponding to the ``armchair'' direction of the graphene hexagonal structure.  We select two boundary strips of atoms at the opposite X-ends of the sheet. These strips are $3.5~\AA$ wide, corresponding to 4~atom layers. Hence, the atoms are free to move  in the Y and Z directions, but follow an imposed motion along the stretching direction (X). This constraint induces an initial pre-stress on the sheet that is visible in the stress-strain curve (see Fig.\ref{fig:examplefailure}b).  The Y-end boundaries are left free.  The system is thermostated by means of  a Berendsen~\cite{Berendsen1984} thermostat with a temperature ranging from 1K to 800K, and a characteristic relaxation time equal to 0.1~ps; the simulation timestep is set to 0.5~fs to insure a correct time integration of the atoms dynamics. These parameters lead to a slightly underdamped atom dynamics. Before the stretching protocol is started, the system is allowed to relax to thermal equilibrium from the initial constrained state. Afterwards, one of the lateral strips is set in motion, so that the sample is subject to a constant engineering strain rate  $\dot{\varepsilon}$ independent of the system size. The  strain rates lie between \(1.28 \times 10^{7} \textrm{s}^{-1}\) and  \(1.28 \times 10^{9} \textrm{s}^{-1}\). As for other molecular dynamics simulations, the simulated strain rates are much higher than those applied experimentally, but
the deformation speed is still much lower than the sound speed in graphene. The chosen strain rate is reached by adiabatically ramping up $\dot{\varepsilon}$,  in order to minimize the creation of shock waves in the material. As a matter of fact, visual inspection of velocity fields shows that shock waves are rapidly damped and do not significantly influence the system dynamics. Simulations are carried on until the graphene sheet fractures. Failure statistics are sampled over 100 realizations for each condition in which we vary vacancy concentration $P$, temperature $T$, strain rate $\dot{\varepsilon}$ and  system size $N$. The only the exception is provided by systems characterized by $T=300$K, $\dot{\varepsilon}=0.128\times 10^8 s^{-1}$,  $N=20\times 10^3$ and $N=50\times 10^3$ atoms, where  50 samples were simulated.

\section{Simulations \label{sec:simulations}}

\begin{figure}
 \centering
 \includegraphics[width=\columnwidth,keepaspectratio=true]{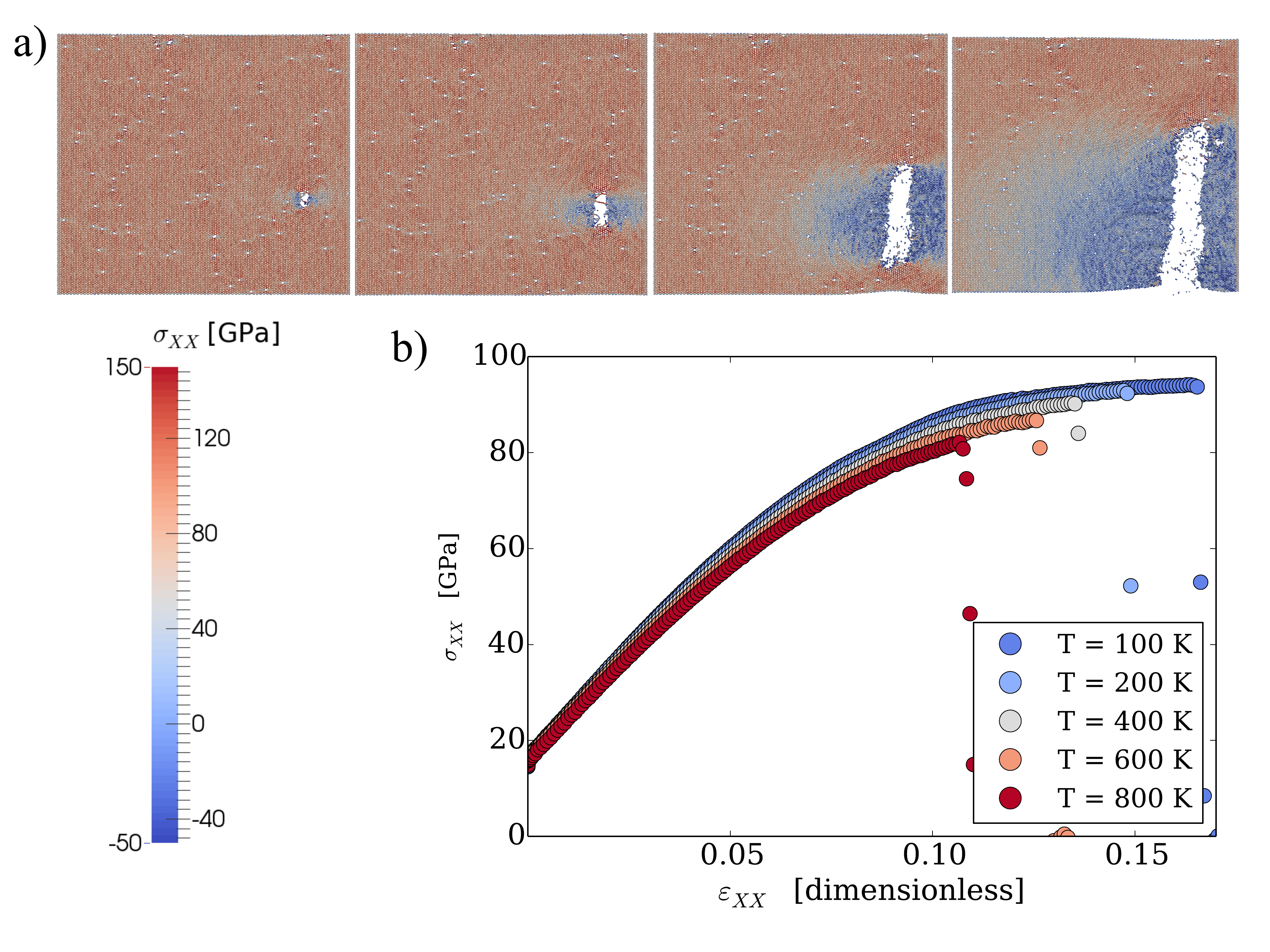}
 \caption{{\bf Failure of graphene sheets.} The graphene sheet is composed of  $N=50\times 10^3$ atoms, with a vacancy concentration (porosity) 
$P=0.1\%$.   The color bar indicates the $\sigma_{xx}$ component of stress tensor per-atom. a) Graphical view of the failure process (from left to right). The crack nucleates from one of the defects already present in the material (not necessarily the most stressed) and rapidly grows untill the graphene sheet complete failure is achieved.
 b)  The stress strain curve displays 
 temperature dependent fracture strength. The pre-stressed initial condition ($\varepsilon=0$) is due to the constraint applied to the atoms belonging to the 4 outmost layers of the sheet, which are subject to the stretching along X.}
 \label{fig:examplefailure}
\end{figure}

\begin{movie}
 \centering
\includegraphics[width=\columnwidth,keepaspectratio=true]{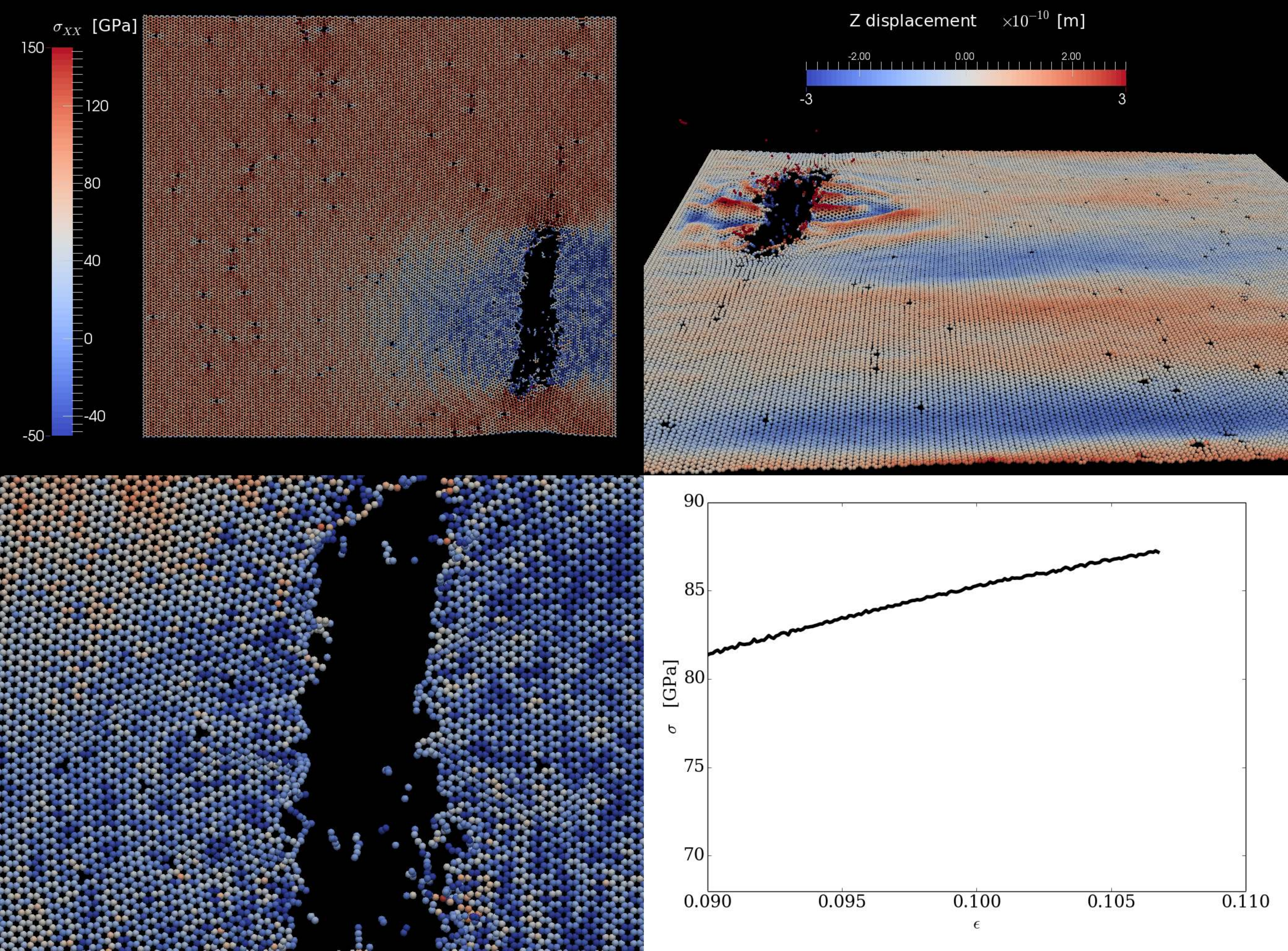}
\caption{The deformation and fracture of a graphene sheet as the strain is ramped is shown in the top left panel ($P=0.1\%$, $N=50\times 10^3$, $T=300$K and $\dot{\varepsilon}=0.128\times 10^8 s^{-1}$). The color represents the tensile stress $\sigma_{XX}$ magnitude. A magnification of the region where the crack is nucleated is shown in the bottom
left panel. The top right panel reports the same sheet viewed under a different angle with a color code representing 
the Z component of the particle positions. The bottom right panel reports the corresponding stress strain curve.}
\label{movie}
\end{movie}

\begin{figure}[ht]
 \centering
 \includegraphics[width=\columnwidth,keepaspectratio=true]{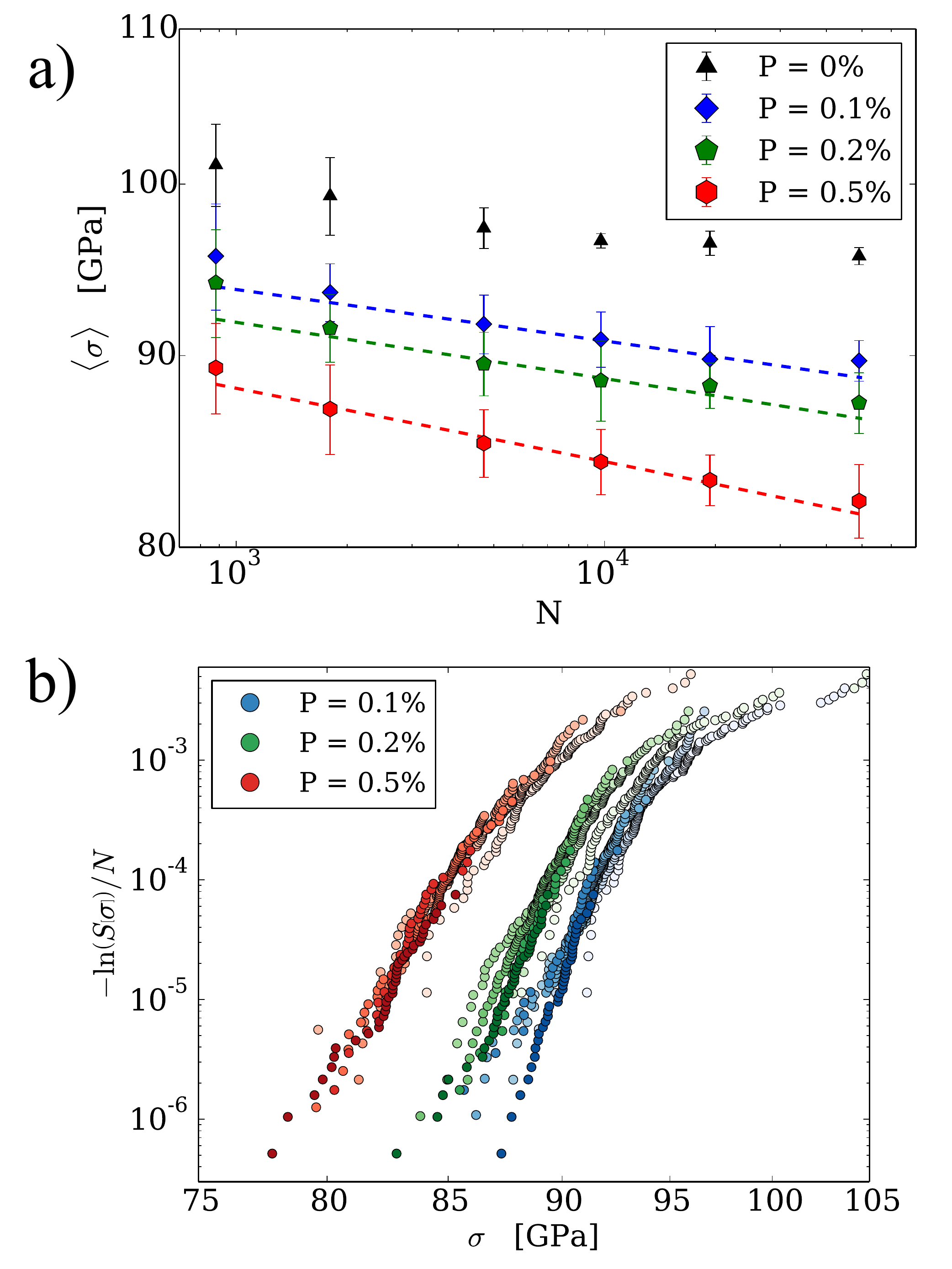}
 \caption{{\bf Graphene fracture size effects.} a) The average failure stress for defected graphene depends on the system $N$ size and on the vacancy concentration $P$. Simulations are carried out with $T=300$K and $\dot{\varepsilon}=0.128\times 10^8 s^{-1}$. The lines are the theoretical prediction as discussed in the supporting information. They do not arise as direct fit of the numerical curves, but result from the analytical evaluation of the integral expression of $\langle\sigma\rangle_n$. b) The failure stress survival distribution at $T=300$K,and $\dot{\varepsilon}=0.128\times 10^8s^{-1} $ for different system sizes with vacancy concentration equal to  $P=0.1$\% (blue) , $P=0.2$\% (green) and $P=0.5$\% (red). When the survival probability distributions are rescaled by $N$ according to the predictions of the extreme value theory, the data collapse into a single curve that only depends on the vacancy concentration $P$.}
\label{fig:distr}
\end{figure}

An example of the fracture process is shown in Fig.~\ref{fig:examplefailure}a, where the graphene structure is seen from above  at four different times during the nucleation and ensuing growth of the crack (see also Video 1). The color code represents the XX component of the symmetric per-atom stress tensor $\sigma_{xx}$, including both potential and kinetic terms. Typical stress strain curves are reported in Fig.~\ref{fig:examplefailure}b, showing that the tensile strength depends on temperature $T$. Our results provide a clear indication that it also depends on system size $N$, vacancy concentration $P$ and strain rate $\dot{\varepsilon}$, as we discuss below.

\begin{figure}[ht]
 \centering
 \includegraphics[width=\columnwidth,keepaspectratio=true]{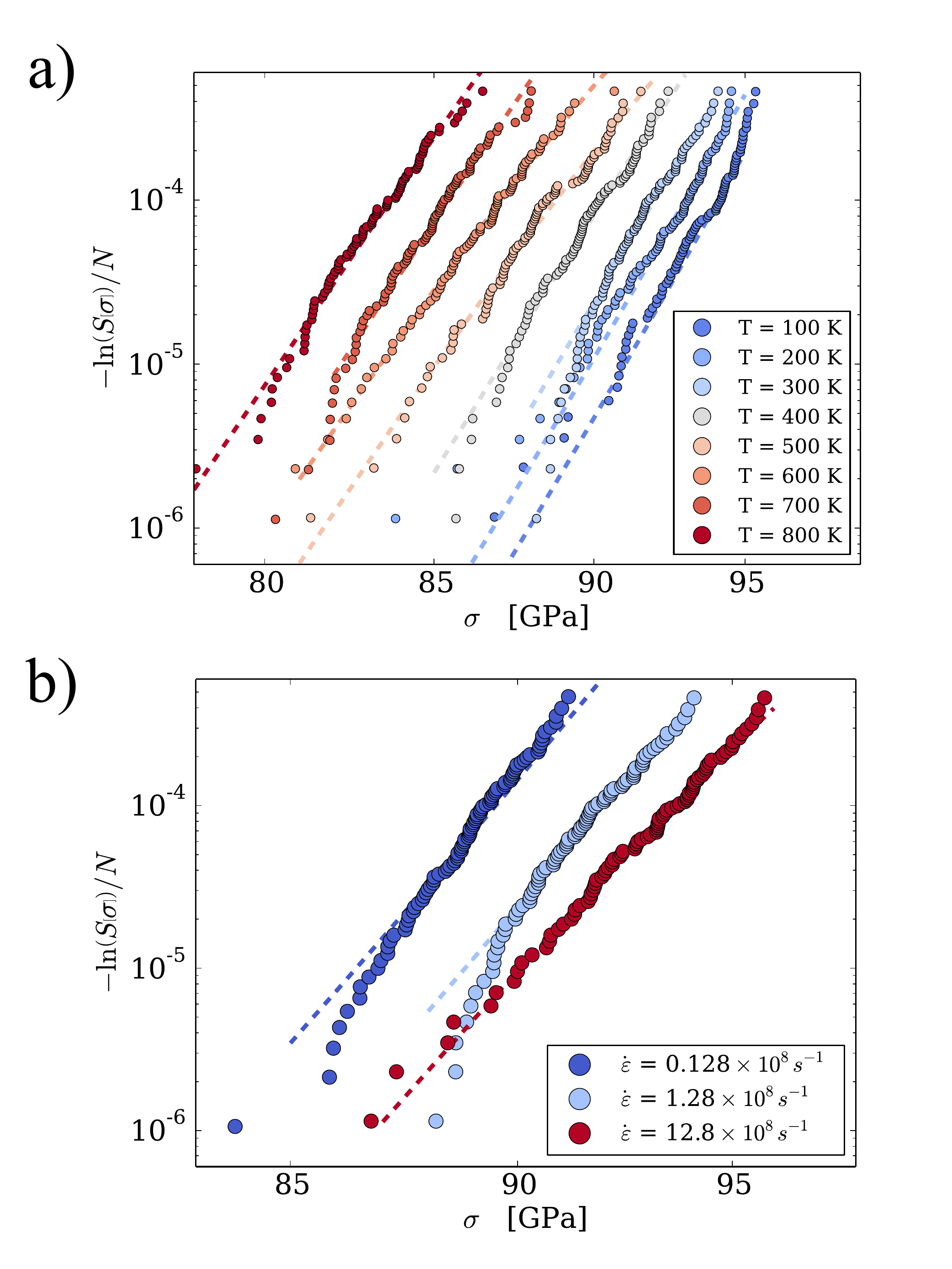}
 \caption{{\bf Temperature and rate effects of the graphene tensile strength distribution.}  
 The survival distribution of defected graphene sheets (with $P=0.2$, $\dot{\varepsilon}=1.28\times 10^8 s^{-1}$ and $N=10^4$) depends on 
 temperature (a) and  strain rate (b: $P=0.2$, $T=300$K and $N=10^4$). The dashed lines represent the best least square fit according to the theory of breaking kinetics discussed in the text.}
\label{fig:distrT-rate}
\end{figure}

\noindent Fig. \ref{fig:distr}a reports the average failure stress $\langle\sigma\rangle$ as a function of system size for different values of the porosity $P$, showing that the larger and more defective a sample is, the weaker it is. A more complete description of the failure
statistics is obtained by the survival distribution $S(\sigma)$, defined as the probability
that a sample has not yet failed at a stress $\sigma$. The numerical results for  $S(\sigma)$
are reported in Fig. \ref{fig:distr}b. If a system of volume $V$ fails according to extreme value statistics, the survival distribution should depend on the volume as $S(\sigma) = S_0(\sigma)^{V/V_0}$, where $S_0(\sigma)$ is the survival distribution of a representative element of volume $V_0$, the smallest independent unit in the sample \cite{alava09}. If we express the volume in terms of the number of atoms $N$ and their atomic volume $V_a$, the survival probability can be written as $S(\sigma)=\exp[- N V_a/V_0 f(\sigma)]$, where $f(x)$ is a suitable function which is a power law $x^\kappa$ in case of   Weibull distribution \cite{weibull39}, and exponential $e^x$ for  Gumbel distribution \cite{gumbel}. Fig. \ref{fig:distr} shows that the $N$-dependence of the survival distribution follow the prescriptions of extreme value theory, but $f(x)$ is not a power law, indicating that the Weibull distribution does not represent the data. This is confirmed by the size scaling of the average failure stress that does not follow a power law, as would be expected from the Weibull distribution. The survival distribution depends also on temperature and strain rate, as shown in Fig. 
\ref{fig:distrT-rate}, which is hard to reconcile with the weakest link hypothesis underlying the Weibull distribution. Indeed, by monitoring
the local stress field $\sigma_{xx}$ before failure, we estimate that only in less than 20\% of the samples (for $N=50\times 10^3$) the final crack nucleates in the most stressed region. In 50-60\% of the cases, the final crack is nucleated in regions that ranked fourth of more in terms of stress. This is a clear indication that failure is not dictated by the weakest link.

\section{Thermal activated fracture of disordered media \label{sec:theory}}

\subsection{Derivation of the survival distribution}
To understand our simulation results, we generalize extreme value theory taking into account thermal fluctuations. 
We describe the system as a set of $n$ representative elements of volume $V_0$ (slabs) such that the thermally activated failure of a single element induces global failure. Each representative $i$ element obeys linear elasticity up to a critical strain $\varepsilon_c^i$, so that the elastic energy of the sample under an external stress $\sigma$ is given by
$U(\sigma) =\sum_i (U_0(\varepsilon_i,\varepsilon_c^i) -V_0\sigma \varepsilon_i)$, where 
\begin{equation}
U_0(\varepsilon,\varepsilon_c)=\left\{
\begin{array}{lll}
V_0\frac{E \varepsilon^2}{2} &   & \varepsilon\leq\varepsilon_c\\ 
-\infty                 &   & \varepsilon > \varepsilon_c,
\end{array}
\right.
\label{potential_0}
\end{equation}
where $E$ is the Young modulus. The sample is loaded at constant strain rate $\dot{\varepsilon}$ so that $\sigma(t)=E\dot{\varepsilon}t$ and critical strains are distributed according to a probability density function $\rho(\varepsilon_c)$. Assuming the slabs noninteracting and identicals, the survival probability for the entire sample is given by the product of the survival probabilities of each representative element $S_n(\sigma|T,\dot{\varepsilon})= \left[S_0(\sigma|T,\dot{\varepsilon})\right]^n$, according to the theory of breaking kinetics \cite{coleman1958strength}. The representative volume survival probability is defined as 
\begin{equation}
S_0(\sigma|T,\dot{\varepsilon})=\int_{\sigma/E}^\infty d\varepsilon_c\rho(\varepsilon_c)\Sigma_0(\sigma|\varepsilon_c,T,\dot{\varepsilon}),
\label{survival_cumulative}
\end{equation}
 where $\Sigma_0\left(\sigma|\varepsilon_c,T,\dot{\varepsilon}\right)$ represents the survival probability of a single slab characterized by a failure strain $\varepsilon_c$. Eq. \ref{survival_cumulative} reduces to the standard extreme value theory when $\Sigma_0\left(\sigma|\varepsilon_c,T,\dot{\varepsilon}\right)=1$, but otherwise depends
on temperature and strain rate. In general, however, the theory predicts that $\log (S_n)/n$ should not depend
on the system size, as verified by our simulations (see Fig. \ref{fig:distr}b).

To estimate the survival distribution of the single slab $\Sigma_0\left(\sigma|\varepsilon_c,T,\dot{\varepsilon}\right)$ we make the phenomenological hypothesis that the material failure arises as a thermally activated process. Historically,  the idea that the solid failure can be described by means of the Kramer's theory, where the intrinsic energy barrier is reduced proportionally to the applied field, has firstly appeared in material science to treat the kinetic fracture of solids under applied stresses,  and dates back to the works of Tobolsky and Eyring \cite{tobolsky1943} and, later, of Zhurkov \cite{zhurkov1965}. More recently it has been successfully  applied  to the study the failure of fibers \cite{phoenix1983statistical}, gels \cite{bonn1998delayed}, wood and fiber-glasses \cite{guarino1999failure}  where the potential energy barrier is given by the Griffith crack nucleation energy \cite{pomeau2002}. 
Most of previous work focused on the thermal dependence of the average strength or
the failure time in creep experiments and did not address the survival distribution and
its size dependence. To this end, we start from recent theories developed for single-molecule pulling, where the molecule rate coefficient for rupture (or unbinding) is modified by the presence of an external time-dependent force
\cite{bell1978models,hummer2003,dudko2003,dudko2006,bullerjahn2014,freund2009,friddle2008,maitra2010model}. 

\noindent In our case, the stress-dependent failure rate
of a single element characterized by a failure strain $\varepsilon_c$, is given by an Arrhenius like form \cite{dudko2006,bullerjahn2014,maitra2010model} 
\begin{equation}
k(\sigma|T,\varepsilon_c)= k_02^{3/2}\left(1-\frac{\sigma}{E\varepsilon_c}\right)e^{\frac{V_0E\varepsilon_c^2}{2k_BT}\left[1-2\left(1-\frac{\sigma}{E\varepsilon_c}\right)^2\right]}
\label{rate}
\end{equation}
 where $k_0$ is the Kramer's escape rate from the potential well described in Eq. \ref{potential_0} \cite{hummer2003,hu2010},
\begin{equation}
k_0=\omega_0\left(\frac{EV_0}{k_BT}\right)^{3/2}\frac{\varepsilon_c}{\sqrt{2\pi}}e^{-\frac{V_0E\varepsilon_c^2}{2k_BT}},
\label{rate_0}
\end{equation}
with a characteristic frequency $\omega_0$. In our numerical simulations one end of the graphene sheet is held fixed, while the other is pulled at  constant strain rate $\dot{\varepsilon}$: this can be interpreted as the action of a stiff  device  \cite{bullerjahn2014,maitra2010model} for which Eq.\ref{rate} has been derived.  $\Sigma_0\left(\sigma|\varepsilon_c,T,\dot{\varepsilon}\right)$ obeys to the following first-order rate equation \cite{freund2009}
\begin{equation}
\frac{d \Sigma_0\left(\sigma|\varepsilon_c,T,\dot{\varepsilon}\right)}{dt}=-k(\sigma(t)|T,\varepsilon_c) \Sigma_0\left(\sigma|\varepsilon_c,T,\dot{\varepsilon}\right),
\label{rate_eq}
\end{equation} where $\sigma(t)=E\dot{\varepsilon}t$. The survival probability is then readily obtained as 
\begin{equation}
\Sigma_0\left(\sigma|\varepsilon_c,T,\dot{\varepsilon}\right)=e^{-\frac{\omega_0}{\dot{\varepsilon}}\sqrt{\frac{V_0E}{\pi k_BT}}\left[e^{-(\varepsilon_c-\sigma/E)^2\frac{V_0E}{k_BT}}-e^{-\varepsilon_c^2\frac{V_0E}{k_BT}}\right]}.
\label{survival_adim}
\end{equation}
Notice that Eq. \ref{survival_adim}, only holds for $\sigma < E\varepsilon_c$ since otherwise the 
element fails with probability one (when $\sigma \simeq E\varepsilon_c$ the Kramer's theory incorrectly predicts $k(\sigma|T,\sigma,\varepsilon_c)\simeq 0$, since  it only holds for energy barriers $\gg k_BT$ \cite{dudko2006}). Finally, inserting  Eq. \ref{survival_adim} in Eq. \ref{survival_cumulative} and, in turns, into the constitutive equation for the theory of breaking kinetics,
we obtain 
\begin{eqnarray}\label{survival_N}
S_n(\sigma|T,\dot{\varepsilon})=
\left(\int_{\sigma/E}^\infty d\varepsilon_c\rho(\varepsilon_c) \times\right.\\ \nonumber
\left.\exp{-\frac{\omega_0}{\dot{\varepsilon}}\sqrt{\frac{V_0E}{\pi k_BT}}\left[e^{-(\varepsilon_c-\sigma/E)^2\frac{V_0E}{k_BT}}-e^{-\varepsilon_c^2\frac{V_0E}{k_BT}}\right]}\right)^n.
\end{eqnarray}

\subsection{Limiting behavior of the theoretical survival distribution}

The survival distribution reported in Eq. \ref{survival_N} is written as a convolution of the disorder distribution $\rho(\varepsilon_c)$ with a temperature and rate dependent
kernel. It is instructive to study its limiting behaviors since this allows to
assess the relevance of thermal and rate dependent effects for fracture statistics.
Our starting point is the expression for the conditional survival probability $\Sigma_0\left(\sigma|\varepsilon_c,T,\dot{\varepsilon}\right)$ reported in Eq. \ref{survival_adim}. It is convenient to study its behavior in term of 
the dimensionless parameter $\lambda\equiv (V_0 E)/(k_B T)$, the ratio between the
elastic energy of a representative volume element and the thermal energy. In terms
of $\lambda$ we can write $\Sigma_0(\lambda)\equiv \exp(-G(\lambda))$, where
\begin{equation}
G(\lambda)=\frac{\omega_0}{\dot{\varepsilon}}\sqrt{\lambda}\left[e^{-(\lambda\varepsilon_c)^2(1-\sigma/(\varepsilon_c E))^2}-e^{-(\lambda\varepsilon_c)^2}\right].
\label{eq:lambda}
\end{equation} 
Thermal fluctuations can be neglected when $G(\lambda) \to 0$, yielding the usual disorder-induced survival probability distribution
\begin{equation}
S_n(\sigma|T,\dot{\varepsilon}) \simeq \left(\int_{\sigma/E}^\infty d\varepsilon_c\rho(\varepsilon_c)\right)^n
\label{suppl:survival_N_highstress}.
\end{equation}

It is interesting to consider first the limit of $\lambda \to \infty$, corresponding to  very low temperature and large representative volume elements. In this limit, the exponential factors in $G(\lambda)$ dominates and the function goes to zero even for small
strain rates. In more generality, thermal fluctuations become negligible when
\begin{equation}
\dot{\varepsilon} \gg \omega_0 \sqrt{\lambda}\left[e^{-(\lambda\varepsilon_c)^2(1-\sigma/(\varepsilon_c E))^2}-e^{-(\lambda\varepsilon_c)^2}\right]. 
\label{eq:thermal}
\end{equation}
Therefore there is a temperature and stress dependent critical strain rate above which we can neglect thermal fluctuations. 

Another interesting limit is the low stress regime (i.e. $\frac{\sigma}{E}\to 0$) where
\begin{equation}
 \Sigma_0\left(\sigma|\varepsilon_c,T,\dot{\varepsilon}\right)\to 1-2\frac{\omega_0}{\dot{\varepsilon}}\sqrt{\frac{E}{V_0}}\left(\frac{V_0}{k_BT}\right)^{3/2}\varepsilon_c\sigma.
\label{suppl:survival_adim_lowstress}
\end{equation}
\noindent Hence, thanks to Eq. \ref{survival_cumulative}, the survival distribution for a representative element is given by
\begin{equation}
 S_0\left(\sigma|T,\dot{\varepsilon}\right)\to 1-2\frac{\omega_0}{\dot{\varepsilon}}\sqrt{\frac{E}{V_0}}\left(\frac{V_0}{k_BT}\right)^{3/2}\langle\varepsilon_c\rangle\sigma
\label{suppl:survival_cumulative_lowstress}
\end{equation}
\noindent where $\langle\varepsilon_c\rangle=\int_0^{\infty}d\varepsilon_c\,\varepsilon_c\rho(\varepsilon_c)$. Therefore, the survival probability distribution function for the entire system can be recast as 
\begin{equation}
-\frac{\ln S_n(\sigma|T,\dot{\varepsilon})}{n}\to2\frac{\omega_0}{\dot{\varepsilon}}\sqrt{\frac{E}{V_0}}\left(\frac{V_0}{k_BT}\right)^{3/2}\langle\varepsilon_c\rangle\sigma
\label{suppl:survival_N_lowstress},
\end{equation}
\noindent displaying a linear dependence on the applied stress, irrespective of the failure strain distribution function $\rho(\varepsilon_c)$.

\subsection{Fit of the numerical data}
\begin{figure}[h]
 \centering
 \includegraphics[width=\columnwidth,keepaspectratio=true]{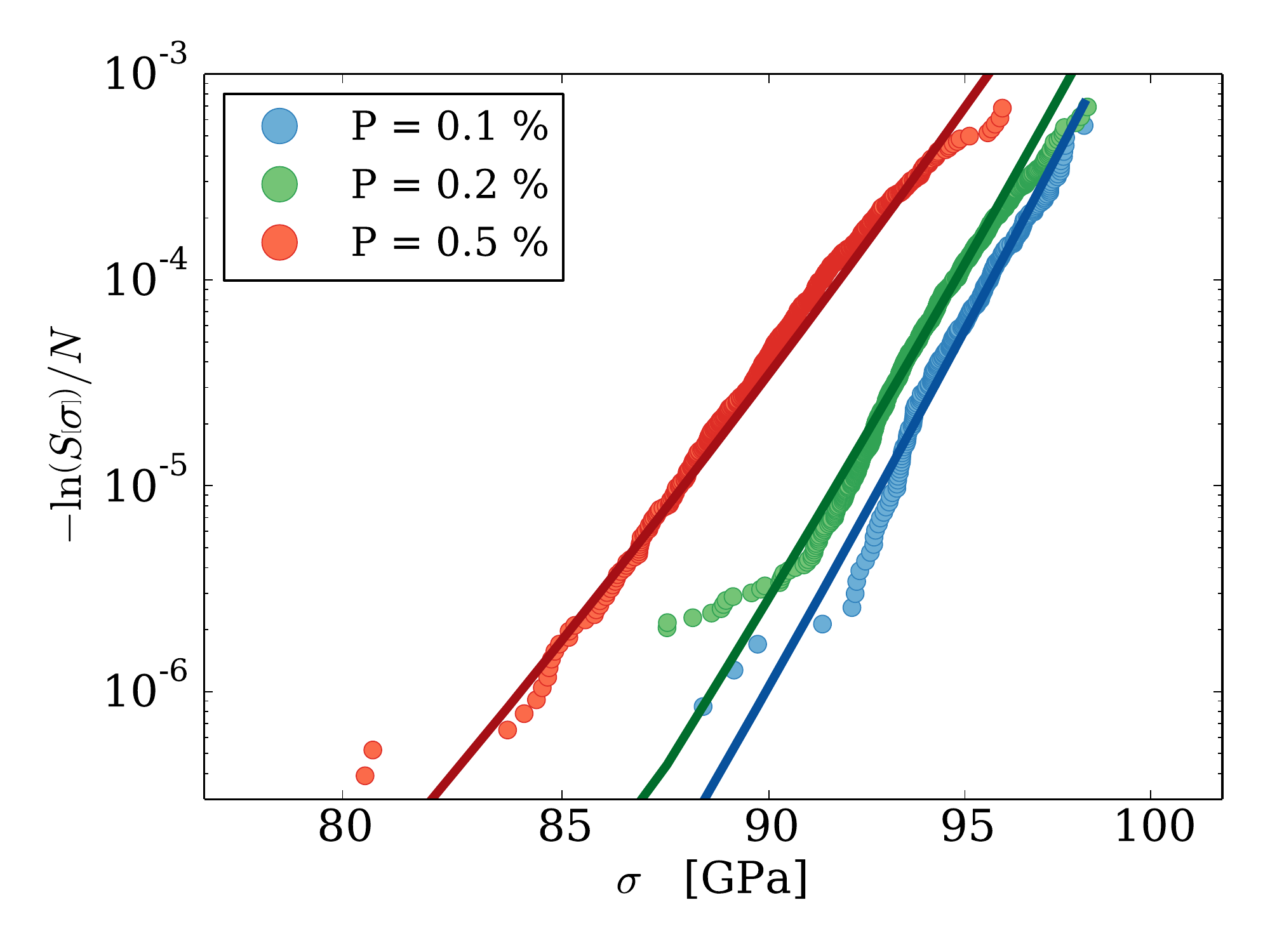}
 \caption{{\bf Survival distribution of defected graphene at low temperature.} We report the survival distribution obtained from  simulations  at $T=1$K, $\dot{\varepsilon}=0.128\times 10^8s^{-1}$ and $N=10^4$ for different values of the vacancy concentration $P$. The numerical data are fitted with
the exponential function $Ae^{\frac{\sigma}{E\varepsilon_0}}$ (solid lines), leading to a  Gumbel distribution for the failure strains $\rho(\varepsilon_c)=Ae^{\frac{\varepsilon_c}{\varepsilon_0}-Ae^{\frac{\varepsilon_c}{\varepsilon_0}}}$ (see Eq.\ref{suppl:rho}). For $P=0.1\%$ we obtain $A=7.92\pm0.05\times 10^{-38}$, $\varepsilon_0=0.00125\pm0.00004$. For $P=0.2\%$: $A=1.767\pm0.005\times 10^{-35}$, $\varepsilon_0=0.001338\pm0.000007$. For $P=0.5\%$ $A=1.804\pm0.007\times 10^{-28}$, $\varepsilon_0=0.00167\pm0.00004$.   }
 \label{fig:lowT}
\end{figure}

\noindent Eq. \ref{survival_N} provides an excellent fit to the results obtained from numerical simulations of defected graphene  at different defect concentrations $P$, temperature $T$ and loading rate $\dot{\varepsilon}$. To fit the numerical simulations with Eq. \ref{suppl:survival_N_highstress}, we first need to establish the form of $\rho(\varepsilon_c)$. This is a phenomenological  function describing the distribution of failure strains of representative volume elements at zero temperature. A reasonable estimate of its functional form can be obtained from simulations at low temperature (i.e. $T=1$K), where thermal fluctuations are negligible, as discussed in details in appendix \ref{sec:fit}. 
The numerical outcomes indicate that $\rho(\varepsilon_c)$ follow the Gumbel distribution \cite{gumbel} (see Fig. \ref{fig:lowT}). We then insert the resulting form of 
$\rho(\varepsilon_c)$ in Eq. \ref{survival_N} which we adopt as a fitting function for the numerical survival probability $S(\sigma)$, with $\omega_0$ and $V_0$ as fitting parameters. 

\begin{figure}[h]
 \centering
 \includegraphics[width=\columnwidth,keepaspectratio=true]{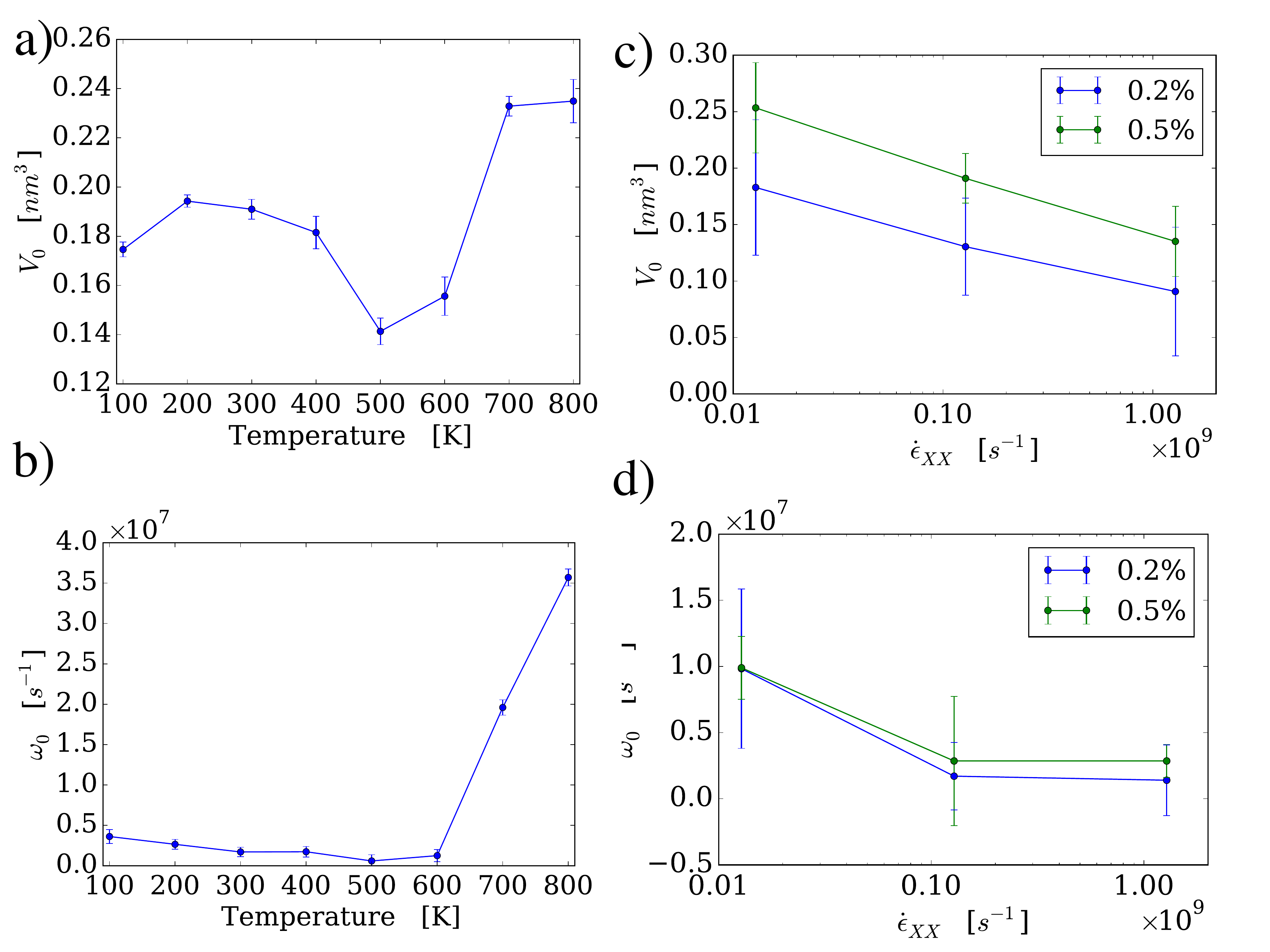}
 \caption{{\bf Fitting parameters for graphene survival distributions.} The best fitted values of
  a) the representative volume $V_0$ and b) the activation frequency as a function of temperature ($N=10^4$, $\dot{\varepsilon}=1.28\times 10^8 s^{-1}$). These values are obtained by the least square fit of the numerical survival probability distribution with the expression \ref{suppl:survival_N} (dashed lines in Fig.3a).  The
 same values, c) and d), as a function of the strain rate ($N=10^4$, $T=300$K), obtained from the best fits shown in Fig.3b ($P=0.2\%$)  and Fig.\ref{fig:P0.5} ($P=0.5\%$) (dashed lines).
 }
 \label{fig:fitting parameters}
\end{figure}

The representative volume $V_0$ ranges between $0.1nm^3$ and $0.25nm^3$, while the characteristic frequency is found in the range $\sim 6\times 10^6 s^{-1}$ and $\sim 10^8 s^{-1}$ (see Fig. \ref{fig:fitting parameters}) Moreover, from the survival distribution we 
also calculate, without additional fitting, the system size dependence of the average tensile strength $\langle\sigma\rangle_n$, which displays an excellent agreement with simulations results as shown in Fig. \ref{fig:distr}a. Further details on the fitting methodology and the analytical expressions used in our model  are reported in appendix \ref{sec:fit}.

\section{Discussion \label{sec:discussion}}

In conclusions, we have performed extensive numerical simulations for the tensile failure
of defected graphene focusing on the size effects of the strength distribution for different
temperatures and loading rate. The results of numerical simulations show deviations from the weakest-link hypothesis but can be explained by taking explicitly into account the effect of thermally activated crack nucleation. The resulting theory describes well our results and could prove useful to understand the tensile strength distribution of other nanomaterials such as carbon nanotubes or other nanowires.

At present it is not possible to compare our numerical and theoretical predictions directly to experiments. Experimental measurements of the strength of graphene sheets are mostly based
on indentation tests \cite{Lee2008}, while tensile tests only recently appeared in the literature \cite{Zhang2014} but thermal, rate and size effects have not been studied. Furthermore, most experimental studies are focusing on the strength of graphene in pristine conditions \cite{Lee2008} without defects or pre-existing cracks. Our theory is, however, very general yielding predictions that  should be applicable also to other carbon nanomaterials and allows to formulate general considerations on the relevance of thermal effects for fracture. 

Eq. \ref{survival_N} suggests that thermal fluctuations can be neglected for
large enough strain rate, since in this limit $\Sigma_0 \simeq 1$ and the sample fails according to the weakest link statistics. In our simulations, we have
$E \simeq 10^{12}$Pa, $V_0 \simeq 0.1\mathrm{nm}^3$ so that at room temperature we estimate 
$\lambda \simeq 10^5$. If we use this value in Eq. \ref{eq:lambda}, we find that 
for $\varepsilon_c \simeq 0.1$ the exponential terms do not vanish close
to failure (i. e. for $\sigma> 0.9 (E\varepsilon_c)$) and thermal effects should therefore be relevant. Indeed using $\frac{\omega_0}{\dot{\varepsilon}} \simeq 10^{-2}$ in 
Eq. \ref{eq:thermal}, one can readily show that thermal effects start to become relevant for $T>10$K in agreement with our simulations.

The same argument suggests that in macroscopic samples, with larger representative volume 
elements, thermally activated failure can often be ignored, even at room temperature. Consider for instance a ceramic material, like sintered $\alpha$-alumina \cite{munro1997},  with $E=10^{11}\mathrm{Pa}$ and a typical tensile strength of $\sigma =10^{8}\mathrm{Pa}$. Assuming that the representative volume element corresponds to a grain size of $V_0 \simeq 1(\mu\mathrm{m})^3$, we can estimate $\lambda \simeq 10^{14}$. Now the exponential factors  
impose that $G(\lambda) \to 0$ even at low strain rates, implying that the strength distribution should be described by conventional extreme value theory. Indeed, experiments show that the strength distribution is described by Weibull statistics with parameters that are largely temperature independent \cite{munro1997}. Our theory thus provides a simple way to estimate  the relevance of thermal and rate dependent effects for fracture. This result could have important implications for applications to micro- and nano-mechanical devices whose reliability may crucially depend on the control of thermally activated failure.

\appendix
\section{Interatomic potential and cutoff tuning \label{sec:potential}}
The carbon-carbon atom interactions were modeled using the ``Adaptive Intermolecular REactive Bond Order'' (AIREBO) potential~\cite{Stuart2000}, which was originally developed as an extension of the ``REactive Bond Order'' potential (REBO)~\cite{brenner2002second}. 
In turn, the REBO potential was developed to describe covalent bond breaking and forming with associated changes in atomic hybridization within a classical potential; it has proven an useful tool for modelling complex chemistry in large many-atom systems. The AIREBO potential improves the REBO potential with an adaptive treatment of non-bonded and dihedral angle interactions that is employed to capture the bond breaking and bond reformation between carbon atom chains.
The analytical form of the AIREBO potential (as discussed in the documentation~\cite{Plimpton1995}) is written as:

$$ E = \frac{1}{2} \sum_{i} \sum_{j \neq i} \left[ E_{ij}^{\textrm{REBO}} + E_{ij}^{\textrm{LJ}} + \sum_{k\neq i,j} \sum_{l\neq i,j,k} E_{ijkl}^{\textrm{Torsion}} \right] $$

The $E^{\textrm{REBO}}$ term has the same functional form as the hydrocarbon REBO potential developed in~\cite{brenner2002second}.
We will not cover here the details of the energetic terms which are thoroughly discussed in the mentioned reference.
In short, the REBO term gives the model its \emph{short to medium} range reactive capabilities, describing short-ranged C-C, C-H and H-H interactions ($r < 2$~\AA). These interactions have strong coordination-dependence through a bond order parameter, which adjusts the attraction between the $i,j$ atoms based on the position of other nearby atoms and thus has 3- and 4-body dependencies. A more detailed discussion of formulas for this part of the potential are given in~\cite{Stuart2000}.
The $E_{ij}^{\textrm{LJ}}$ term adds longer-ranged interactions ($2 < r < r_{\textrm{cutoff}}$ \AA) using a form similar to the standard Lennard-Jones potential. It contains a series of switching functions so that the short-ranged LJ repulsion ($1/r^{12}$) does not interfere with the energetics captured by the $E_{ij}^{\textrm{REBO}}$ term. The extent of the $E_{ij}^{\textrm{LJ}}$ interactions is determined by a cutoff argument; in general the resulting $E_{ij}^{\textrm{LJ}}$ cutoff is approximately 10~\AA, in this work we consider a cutoff of approximately 14~\AA.
Finally, the $E_{ijkl}^{\textrm{Torsion}}$ term is an explicit 4-body potential that describes various dihedral angle preferences in hydrocarbon configurations. 

The AIREBO potential has been extensively used to simulate and predict mechanical properties of carbon-based materials, i.e. fullerene, carbon nanotube and graphene~\cite{Zhao2009}. Furthermore, it offers a valid tradeoff between accuracy and computational efficiency; a realistic fracture of large system sizes can be simulated in reasonably short time scales (a few hours on recent computers).
Other interaction models can offer little improvement to the actual realism of the simulation, at the cost of much larger computational costs: for example, the ReaxFF potential, or DFT semiclassical approaches could describe more accurately the fast time scales of chemical reactions, but this would not change the ultimate failure length of the C-C bond: the expected maximum elongation for a C-C bond in graphene is around 0.178~nm.
On the other hand, the use of faster but too simplistic models (e.g. Lennard-Jones potentials, mass and spring systems, or other elastic models) fail to significantly reproduce a realistic behavior.

However, in order to simulate a realistic bond failure behavior, the short-scale C-C adaptive cutoff ($r_c$) of the AIREBO potential has to be tuned.
In fact, it has been observed~\cite{belytschko2002atomistic, zhang2013effect} that, during simulations of fracture of covalent bonds and without cutoff tuning, the shortest-scale potential introduces a sharp increase of bond forces near the cutoff distances, which in turn causes spurious increase in fracture stress and strain~\cite{Zhao2009}. It should also be noted here that this phenomenon is specifically relevant for perfect graphene and CNT lattices, while it is much less pronounced in defected samples, due to the disorder induced in the lattice by the atom vacancies.
This issue has been solved in the past by incrementing the short-scale cutoff lenght of the potential;
the cited papers increase this parameter to 2.0~\AA.
This, however, has the side effect of leading to a singular behavior in the atomic pair potential when the atom atom distance is exactly 2~\AA.

\begin{figure}[htb]
 \centering
 \includegraphics[width=\columnwidth,keepaspectratio=true]{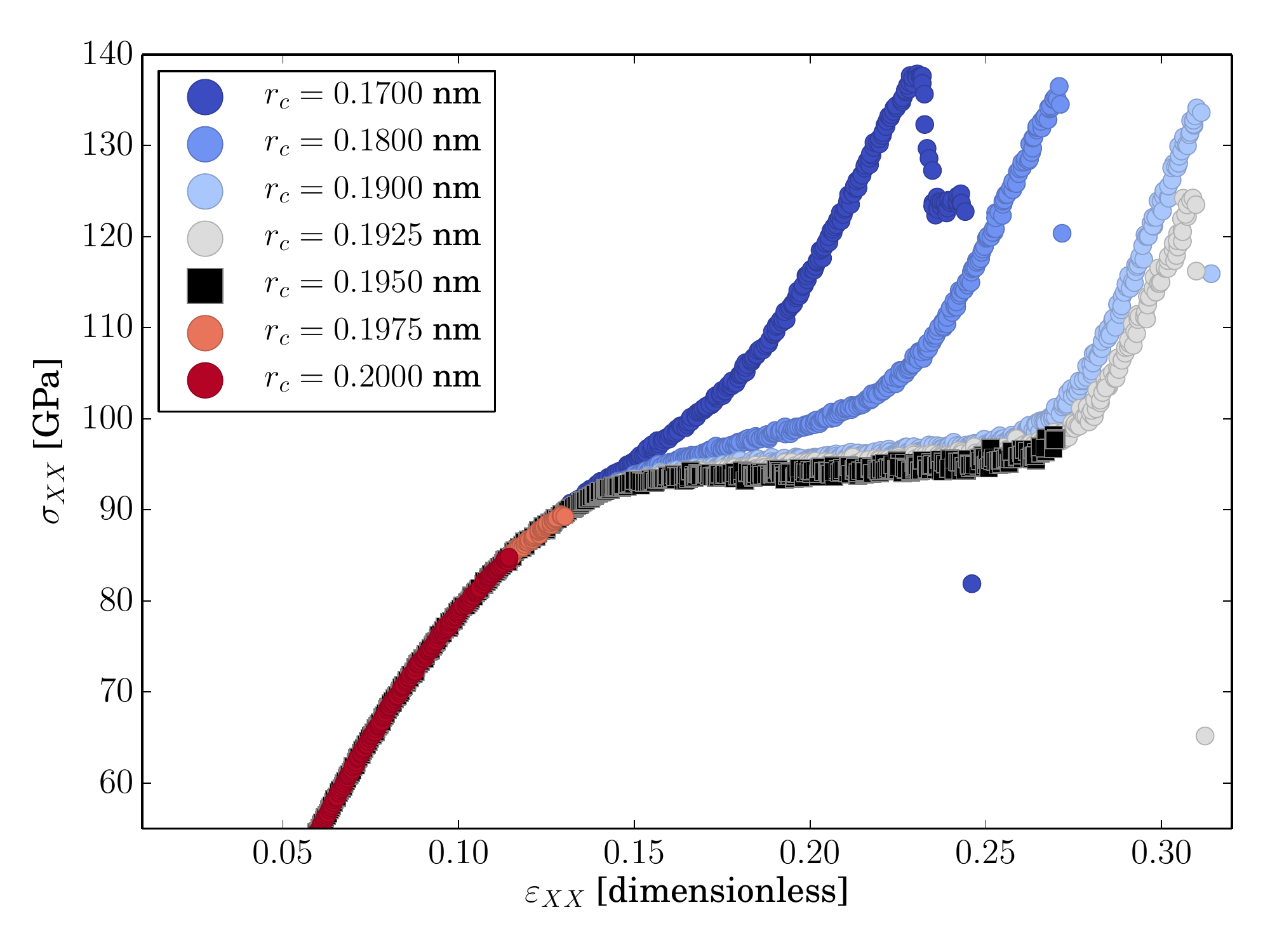}
 \caption{{\bf Tuning the AIREBO potential.} The stress-strain curve obtained as a function of the 
 cutoff $r_c$. The simulated graphene sheet is composed by $N=10^4$ carbon  atoms at $T=300$K, without in-built defects ($P=0$) and pulled at constant strain rate $\dot{\varepsilon}=0.128\times 10^8 s^{-1}$. For $r_c < 0.195$nm, the stress displays a spurious increase while for $r_c=2$nm the pair potential shows an unphysical singularity (not shown). The chosen value of $r_c$ is set to $0.195$nm.}
 \label{fig:cutoff}
\end{figure}

We performed stretching simulations varying the cutoff parameter from $r_c=0.17$~nm (default value) to $r_c=0.2$~nm in both armchair (X) and zigzag (Y) directions of the graphene sheet with no vacancies ($P=0$). The stress-strain curve obtained from the numerical simulations are shown in Fig. \ref{fig:cutoff}. For $r_c<0.195$~nm, a sharp increase on tensile stress for large strains is observed, leading to an unphysical ultra-high failure stress and corresponding failure strain. Increasing the $r_c$ in the range $0.195\leq r_c\leq 0.2$~nm  strongly suppresses this phenomenon. 
Moreover, the stress-strain data reported in Fig. \ref{fig:cutoff}, clearly display  that the failure strain varies from 0.13 to 0.25  when $r_c$ is  in the range $1.95 < r_c \le 2$, whereas the failure stress exhibites a much weaker fluctuation (from $85\times 10^9$ Pa to $95\times 10^9$ Pa). Finally we notice that for defected samples like those investigated in the present article, i.e. $P\neq 0$, the values of the failure stresses and strains do show a much less marked dependence on the choice of $r_c$, whenever $1.95 < r_c \le 2$.

\section{Details of the fitting method \label{sec:fit}}
\noindent 
To fit the numerical simulations with Eq. \ref{survival_N}, we first obtain 
$\rho(\varepsilon_c)$ from simulations at low temperature (i.e. $T=1$K).
As shown in Fig.\ref{fig:lowT}, the numerical survival distribution function $-\frac{\ln S(\sigma)}{N}$, obtained at $T=1$K and $\dot{\varepsilon}=0.128\times 10^8 s^{-1}$, can be nicely fitted  with the following exponential form $Ae^{-\frac{\sigma}{E\varepsilon_0}}$. The theoretical prediction for the survival probability distribution furnished by Eq.\ref{suppl:survival_N_highstress} requires $-\ln \int_{\sigma/E}^\infty d\varepsilon_c\rho(\varepsilon_c)=Ae^{-\frac{\sigma}{E\varepsilon_0}}$, once we assume that $V_0\equiv V_a$ when $T\to 0$. Hence,  we obtain 
\begin{equation}
\rho(\varepsilon_c)=Ae^{\frac{\varepsilon_c}{\varepsilon_0}-Ae^{\frac{\varepsilon_c}{\varepsilon_0}}}
\label{suppl:rho},
\end{equation}
\noindent which corresponds to a Gumbel distribution of failure strains \cite{gumbel}. The numerical values of the fitting parameters $A,\varepsilon_0 $ are reported in the caption of Fig.\ref{fig:lowT} for three vacancy concentrations $P$. We notice that the simulated samples for $T=1$K are 250 in the case of $P=0.1\%$ and 850 for $P=0.2\%$ and 800 $P=0.5\%$.

\begin{figure}[htb]
 \centering
 \includegraphics[width=\columnwidth,keepaspectratio=true]{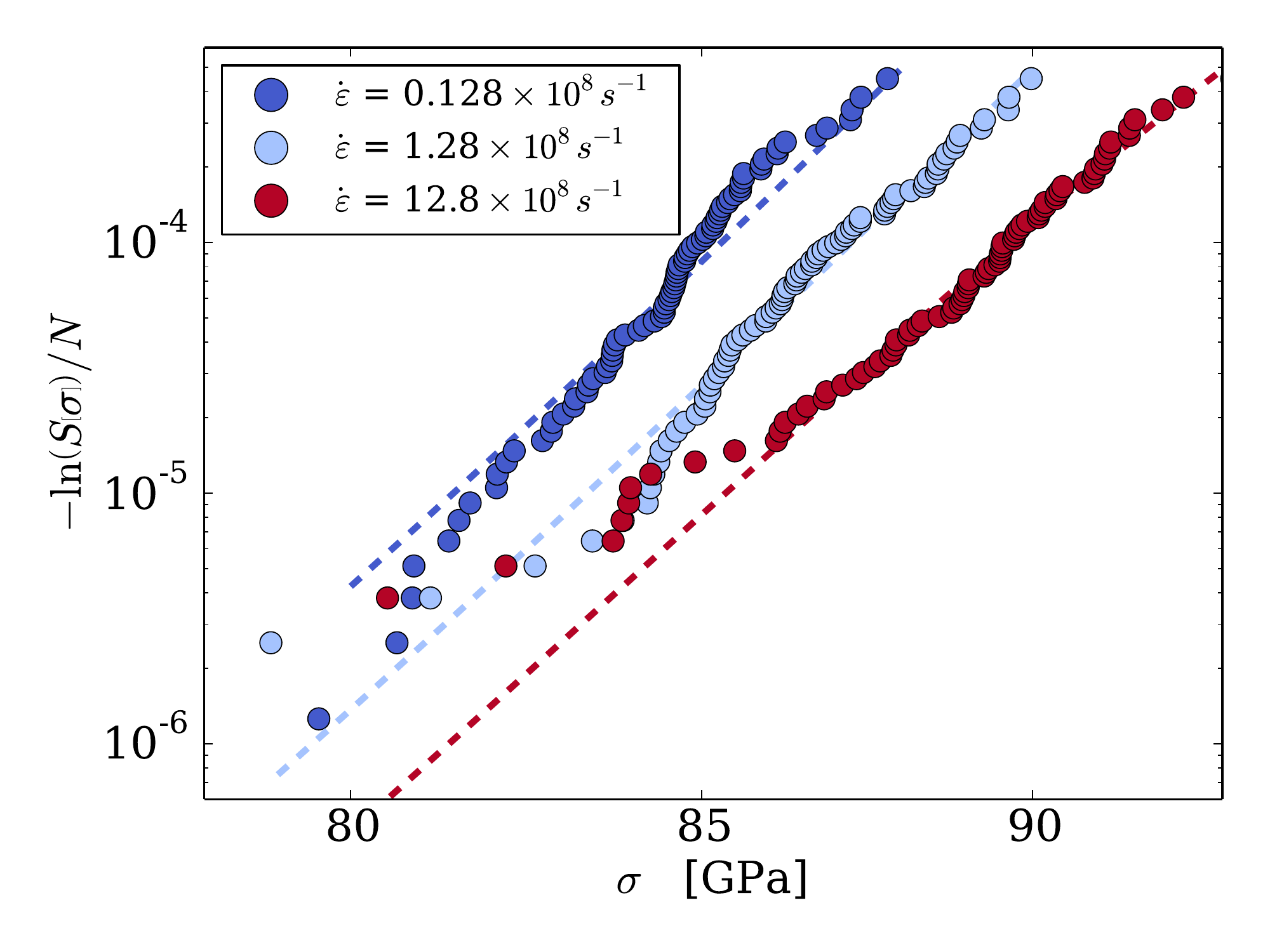}
 \caption{{\bf Survival distribution of defected graphene at P=0.5\%.} We report the survival distribution obtained from  simulations  at $T=300\mathrm{K}$, $N=10^4$ and three strain rates $\dot{\varepsilon}$. The vacancy concentration is set to $P=0.5\%$. The fitting function is provided by Eq.\ref{suppl:survival_N}: the values of the fitting parameters $V_0$ and $\omega_0$ are reported in Fig.\ref{fig:fitting parameters}.}
 \label{fig:P0.5}
\end{figure}

\begin{figure}[htb]
 \centering
 \includegraphics[width=\columnwidth,keepaspectratio=true]{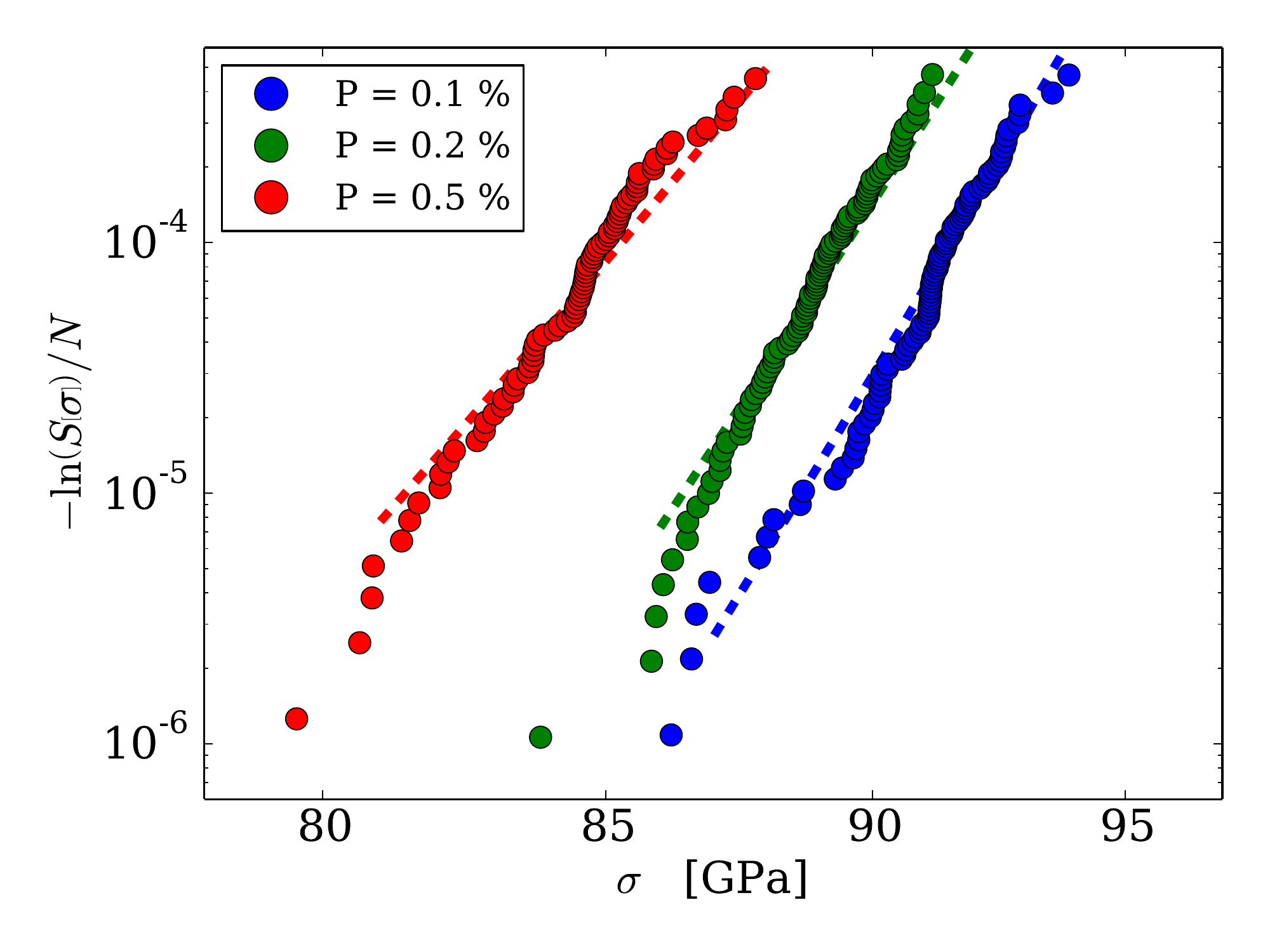}
 \caption{{\bf Survival distribution of defected graphene at different P} We report the survival distribution obtained from  simulations  at $T=300\mathrm{K}$, $\dot{\varepsilon}=0.128\times 10^8 s^{-1}$ and $N=10^4$ for three vacancy concentrations $P$. The numerical data are fitted with the expression \ref{suppl:survival_N} (dashed lines) using the proper values of $A$ and $\varepsilon_0$ reported in Fig.\ref{fig:lowT}. The fitted values of $V_0$ and $\omega_0$ are shown in  Fig.\ref{fig:fitting parameters} for $P=0.2\%$ and $P=0.5\%$. For $P=0.1\%$ the least square fit gives $V_0=0.3806\pm 0.0003 nm^3$ and $\omega_0=4.1416\pm0.0006\times 10^7 s^{-1}$. The set of parameters $A, \varepsilon_0, V_0$ and $\omega_0$ which characterize uniquely the theoretical expression \ref{suppl:survival_N} (dashed lines) are  inserted  into Eq.\ref{suppl:mean_stress} to calculate the mean average rupture stress $\langle\sigma\rangle_n$ as a function of $N$, shown in Fig.2a.}
 \label{fig:various_P}
\end{figure}

\noindent We then perform the least square fit of the numerical  survival probabilities $-\frac{\ln S(\sigma)}{N}$, obtained for different values of $T$, $\dot{\varepsilon}$ and porosities $P$ (see Fig.s 3a,b, \ref{fig:P0.5}, \ref{fig:various_P}), with the following function 
\begin{widetext}
\begin{equation}
-\frac{\ln S_n(\sigma|T,\dot{\varepsilon})}{N}=-\frac{V_a}{V_0}\ln\left(\int_{\sigma/E}^\infty d\varepsilon_c Ae^{\frac{\varepsilon_c}{\varepsilon_0}-Ae^{\frac{\varepsilon_c}{\varepsilon_0}}} e^{-\frac{\omega_0}{\dot{\varepsilon}}\sqrt{\frac{V_0E}{\pi k_BT}}\left[e^{-(\varepsilon_c-\sigma/E)^2\frac{V_0E}{k_BT}}-e^{-\varepsilon_c^2\frac{V_0E}{k_BT}}\right]}\right),
\label{suppl:survival_N}
\end{equation}
\end{widetext}
\noindent where the fitting parameters are the representative  volume element $V_0$ and the characteristic frequency $\omega_0$. The atomic volume $V_a$ has been evaluated by considering a density of 38.18 atoms per $nm^2$ and a sheet thickness equal to $0.335$ nm, yielding $V_a=8.744 \times 10^{-3} nm^3$. For any value of $P$, the corresponding values of $A$ and $\varepsilon_0 $ obtained from the best fit of the data in  Fig.\ref{fig:lowT} are plugged into Eq.\ref{suppl:survival_N}. The fitted $V_0$ and $\omega_0$ corresponding to Fig.s 3a,b, \ref{fig:P0.5}, \ref{fig:various_P} are reported in Fig.\ref{fig:fitting parameters}.

Finally we provide the analytical expression for the distribution of failure stresses defined as $P_n\left(\sigma|T,\dot{\varepsilon}\right)=-\frac{dS_n}{d\sigma}$:
\begin{widetext}
\begin{equation}
\begin{array}{c}
P_n\left(\sigma|T,\dot{\varepsilon}\right)=n\frac{S_{n-1}\left(\sigma|T,\dot{\varepsilon}\right)}{E}\left\{Ae^{\frac{\sigma}{E\varepsilon_0}-Ae^{\frac{\sigma}{E\varepsilon_0}}}e^{-\frac{\omega_0}{\dot{\varepsilon}}\sqrt{\frac{V_0E}{\pi k_BT}}\left[1-e^{-\frac{V_0\sigma^2}{Ek_BT}}\right]}+\right.\\
+\left.\frac{\omega_0}{\sqrt{\pi}\dot{\varepsilon}}\left(\frac{V_0E}{ k_BT}\right)^{3/2}\int_{\sigma/E}^\infty d\varepsilon_c\,Ae^{\frac{\varepsilon_c}{\varepsilon_0}-Ae^{\frac{\varepsilon_c}{\varepsilon_0}}}\left(\varepsilon_c-\frac{\sigma}{E}\right)e^{-(\varepsilon_c-\sigma/E)^2\frac{V_0E}{k_BT}}e^{-\frac{\omega_0}{\dot{\varepsilon}}\sqrt{\frac{V_0E}{\pi k_BT}}\left[e^{-(\varepsilon_c-\sigma/E)^2\frac{V_0E}{k_BT}}-e^{-\varepsilon_c^2\frac{V_0E}{k_BT}}\right]}\right\}.
\end{array}
\label{suppl:probability_N_final}
\end{equation}
\end{widetext}
\noindent Eq.\ref{suppl:probability_N_final} allows to derive the mean failure stress as \pagebreak

\begin{widetext}
\begin{equation}
\begin{array}{c}
\langle\sigma\rangle_n=n\int_0^\infty d\sigma\,\frac{\sigma}{E}S_{n-1}\left(\sigma|T,\dot{\varepsilon}\right)\left\{Ae^{\frac{\sigma}{E\varepsilon_0}-Ae^{\frac{\sigma}{E\varepsilon_0}}}e^{-\frac{\omega_0}{\dot{\varepsilon}}\sqrt{\frac{V_0E}{\pi k_BT}}\left[1-e^{-\frac{V_0\sigma^2}{Ek_BT}}\right]}+\right.\\
+\left.\frac{\omega_0}{\sqrt{\pi}\dot{\varepsilon}}\left(\frac{V_0E}{ k_BT}\right)^{3/2}\int_{\sigma/E}^\infty d\varepsilon_c\,Ae^{\frac{\varepsilon_c}{\varepsilon_0}-Ae^{\frac{\varepsilon_c}{\varepsilon_0}}}\left(\varepsilon_c-\frac{\sigma}{E}\right)e^{-(\varepsilon_c-\sigma/E)^2\frac{V_0E}{k_BT}}e^{-\frac{\omega_0}{\dot{\varepsilon}}\sqrt{\frac{V_0E}{\pi k_BT}}\left[e^{-(\varepsilon_c-\sigma/E)^2\frac{V_0E}{k_BT}}-e^{-\varepsilon_c^2\frac{V_0E}{k_BT}}\right]}\right\}.
\end{array}.
\label{suppl:mean_stress}
\end{equation}
\end{widetext}
\noindent This quantity can be analytically calculated and plotted as a function  of $N$, setting $n=N\frac{Va}{V_0}$, as shown in Fig.2a for $T=300$K, $\dot{\varepsilon}=0.128\times 10^8 s^{-1}$ and  three values of the vacancy concentration $P$. We emphasize that in this case no fit, but just the numerical evaluation of the integral expression of $\langle\sigma\rangle_n$ \ref{suppl:mean_stress} is provided, making use of the proper values of $A, \varepsilon_0, V_0, \omega_0$, obtained by fitting  the  survival probabilities displayed in Fig.\ref{fig:lowT} and Fig.\ref{fig:various_P}.

\clearpage
\acknowledgments
This work is supported by the European Research Council through the Advanced Grant No 29001 SIZEFFECTS.



%

\end{document}